\documentclass{aa}
\usepackage{times,psfig}

\def\rxs{1RXS\,J170849.0--400910}
\def\ltsima{$\; \buildrel < \over \sim \;$}
\def\lsim{\lower.5ex\hbox{\ltsima}}
\def\gtsima{$\; \buildrel > \over \sim \;$}
\def\gsim{\lower.5ex\hbox{\gtsima}}

\newcommand{\be}{\begin{equation}}
\newcommand{\en}{\end{equation}}
\newcommand{\ergs}{\rm \ erg \; s^{-1}}

\def\cmdue {\rm \ cm^{-2}}

\def\deg {^\circ}

\begin{document}

\title{{\em Swift} and {\em Chandra} confirm the intensity-hardness correlation of the AXP
1RXS\,J170849.0--400910}

\author{S.~Campana\inst{1}, N.~Rea\inst{2}, G.L.~Israel\inst{3},
R.~Turolla\inst{4}, S.~Zane\inst{5}}

\authorrunning{S. Campana}

\titlerunning{{\em Swift} and {\em Chandra} observations of 1RXS J170849.0--400910}

\offprints{Sergio Campana, campana@merate.mi.astro.it}

\institute{INAF-Osservatorio Astronomico di Brera, Via Bianchi 46, I--23807
Merate (Lc), Italy
\and
SRON - Netherlands Institute for Space Research, Sorbonnelaan 2, 3584 CA,
Utrecht, the Netherlands
\and
INAF-Osservatorio Astronomico di Roma,
Via Frascati 33, I--00040 Monteporzio Catone (Roma), Italy
\and
Department of Physics, University of Padua, via Marzolo 8, I-35131, Padova, Italy
\and
Mullard Space Science Laboratory, University College London, Holmbury St.
Mary,
Dorking Surrey, RH5 6NT, UK
}

\date{received, accepted}

\abstract{Convincing evidence for long-term variations in the emission
properties of the anomalous X-ray pulsar \rxs\ has been gathered in the last
few years. In particular, and following the pulsar glitches of 1999 and 2001,
{\em XMM-Newton} witnessed in 2003 a decline of the X-ray flux accompanied by
a definite spectral softening. This suggested the existence of a correlation
between the luminosity and the spectral hardness in this source, similar to
that seen in the soft $\gamma$-repeater SGR 1806--20.  
Here we report on new {\em Chandra} and {\em Swift}
observations of \rxs\ performed in 2004 and 2005, respectively. These
observations confirm and strengthen the proposed correlation. The trend
appears to have now reversed: the flux increased and the spectrum is now harder. 
The consequences of these observations for the twisted magnetosphere scenario 
for anomalous X-ray pulsars are briefly discussed. 
}

\maketitle

\section{Introduction}
\label{intro}

The Anomalous X--ray Pulsars (AXPs) are a small group of sources
which stand apart from other known classes of X-ray pulsars. In
particular, they all rotate with spin periods clustered in a very
narrow range ($\mathrm P\sim 5-12$\,s), they have large period derivatives
($\dot{\mathrm P}\sim 10^{-13} - 10^{-10}$\,s\,s$^{-1}$), and, except in one
case (Camilo et al.~2006), deep searches for radio pulsations gave so far
always negative results (Burgay et al.~2006). Another important characteristic
which motivated the  ``anomalous'' label (Mereghetti \& Stella 1995; van Paradijs, Taam
\& van den Heuvel 1995), is their relatively high X-ray luminosity
($\sim 10^{34}-10^{36}\ergs$), which cannot be accounted for by
rotational energy losses alone, and no convincing evidence for a companion
star was discovered so far for any of them. These considerations
quite naturally led to the idea that a non-standard energy
production mechanism is involved in their emission.

Many different models have been suggested all along for AXPs, such as they are
accreting from a fossil disk, formed by the debris of the supernova event, or
from a very low-mass companion (e.g. Mereghetti \& Stella 1995; Mereghetti et
al. 1998; Chatterjee et al. 2000; Perna et  al.~2000; Alpar~2001). 
On the other hand, many observational properties support the idea of these
sources being magnetars, i.e. isolated neutron stars powered by the decay of 
their huge magnetic fields ($\mathrm B\sim10^{14}-10^{15}$\,G; Duncan \&
Thompson~1992; Thompson \& Duncan 1993, 1995, 1996). In fact, if the large
observed spin-down is interpreted in terms of magneto-dipolar losses, all the
AXPs seem to have magnetic fields in excess of the quantum critical field 
($\mathrm B>4.4\times10^{13}$\,G). If this is the case,
AXPs should be related to the Soft $\gamma-$ray Repeaters (SGRs), another class of
X-ray sources thought to involve strongly magnetic neutron stars (see Woods \&
Thompson 2004 for a recent review on SGRs/AXPs). In recent years, intense
monitoring programs revealed several common features between AXPs and SGRs,
(i.e. short bursts, weak IR counterparts, high energy tails; Gavriil, Kaspi \&
Woods 2002; Kaspi et al. 2003; Israel et al. 2003; Kuiper, Hermsen \& M\'endez
2004), strengthening the idea of an underlying relation between these two
classes of sources. 

AXPs' spectra in the X-ray range are well described by an empirical model,
made by an absorbed black body ($\mathrm {kT}\sim 0.3-0.6$ keV) plus a
relatively steep power law with (photon index $\Gamma\sim 2-4$), and a hard
X-ray power-law tail with $\Gamma\sim1$. Until a few years ago AXPs were 
commonly believed to be steady X-ray emitters (even if hints for variability
were already found, see Iwasawa et al. 1992; Baykal \& Swank 1996; Oosterbroek
et al. 1998) but recently flux changes and spectral variability were detected,
both long-term and with spin phase (Kaspi et al.~2003; Mereghetti et al.~2004;
Rea et al.~2005). 

1RXS J170849.0--400910 is a prototypical AXP, with a period of $\sim11$\,s
(Sugizaki et al.~1997; Israel et al. 1999), a spin-down rate of
$\sim2\times10^{-11}$s\,s$^{-1}$, and a soft spectrum (Israel et
al. 2001). A phase-coherent timing solution, inferred thanks to the long
{\em Rossi-XTE} monitoring of this source, led to the discovery
of two glitches in the last few years, with very different
post-glitch behavior (Kaspi, Lackey \& Chakrabarty 2000; Dall'Osso et
al. 2003; Kaspi \& Gavriil 2003). In a very recent paper, Rea et
al.~(2005) showed that both the flux and spectral hardness
reached a maximum level close to the two glitches that the source
experienced in 1999 and 2001, and then decreased again in close
correlation. Moreover, a long observation taken by {\em BeppoSAX}
during the recovery from the second, more dramatic, glitch revealed
evidence for a relatively broad absorption line at $\sim 8$ keV (Rea et
al.~2003), not re-detected in more recent pointings (Rea et al.~2005). The 
spectral feature has been interpreted as a cyclotron resonance
feature, yielding an estimate of the neutron star magnetic
field of either $9.2\times10^{11}$ G or $1.6\times10^{15}$ G, in the
case of electron or proton cyclotron absorption, respectively (Rea et
al.~2003).
In this paper we present new {\em Swift}  (calibration) and {\em Chandra} 
observations of {\rxs}. In Sect. \ref{data} and \ref{chandra} we describe the
timing and spectral analysis of the {\em Swift} and {\em Chandra}
observations, respectively, as well as reporting the results. Discussion
follows in Sect. \ref{disc}. 

\section{{\em Swift} observations}
\label{data}

\rxs\, was observed with the {\em Swift} satellite (Gehrels et al.~2004) a few
times, being a calibrator for timing accuracy and for the wings of the
Point Spread Function of the X--Ray Telescope (XRT, Burrows et
al.~2005; see Table \ref{obslog} for a detailed log of the
observations).  Here we focus on data taken in Window Timing (WT) mode
and in Photon Counting (PC) mode\footnote{{\em Swift}--XRT can observe
sources in three observing modes: Low Rate Photodiode (LRPD), Window
Timing (WT) and Photon Counting (PC), having a timing resolution of 0.14 ms,
1.8 ms, 2.5 s, respectively. Each mode is designed to deal 
with sources of different intensities in order to minimize the effects
of photon pile-up but losing spatial information. In LRPD the entire
CCD is read as a photodiode and there is no spatial information. In WT
mode a 1D image is obtained reading data compressing along the central
200 pixels in a single row. PC data produce standard 2D images. For
more details see Hill et al. (2005).} longer than 1 ks.  In
particular, we used the PC data for spectral analysis and the PC + WT
data for timing analysis. This choice is dictated by the fact that WT
observations are affected by the source being at the edge of the
window and during part of the observation the source fell outside the
window, making difficult a secure evaluation of the instrument
spectral response.

%%%%%%%%%%%%%%%%%%%%%%%%%%%%%%%%%%%%%

\begin{table}
\begin{center}
\caption{{\em Swift} observations of \rxs $^{a}$. }
\label{obslog}
\begin{tabular}{cccc}
\hline
\hline
Obs. number & Mode & Exp. time (ks) & Start time\\
\hline
00050700001 & LRPD     & 0.3 & 2005-01-29\\
            &{\it WT }  & {\it 10.3} & \\
            & PC       & 0.7  &\\
00050700002 & WT       & 0.2 & 2005-01-28\\
00050700006 & WT       & 0.2 & 2005-03-23\\
            &{\bf  PC}   & {\bf 2.3}& \\
00050700256 & WT       & 0.6 & 2005-03-20\\
            & PC       & 0.1 & \\
%00050700257 & WT       & 0.2 & \\
%            & PC       & 0.1 & \\
00050701001 & LRPD     & 1.8 & 2005-01-30\\
            & {\it WT}  & {\it 16.5}& \\
            & {\bf PC}  & {\bf 2.3} &\\
00050701002 & WT       & 0.7 & 2005-02-24\\
            & {\bf PC}  & {\bf 10.5}& \\
00050702001 & {\bf PC}  & {\bf 4.6} & 2005-02-02\\
00050702002 & {\bf PC}  & {\bf 1.7} & 2005-02-23\\
00050702003 & PC       & 0.7 & 2005-03-23\\
00050702004 & PC       & 0.5 & 2005-03-29\\
\hline
\hline
\end{tabular}
\begin{flushleft}
{\small $a$ -- Observations in bold face are the ones used for spectral and
timing analysis. Observation in italic face have been used for timing analysis
only. Observations shorter than 100 s were not listed above.}

\end{flushleft}
\end{center}
\end{table}

%%%%%%%%%%%%%%%%%%%%%%%%%%%%%%%%%%%%%%%%%%%%%%%%%%%%%%%%%%%

Data were analysed with the FTOOL task {\tt xrtpipeline} (version
build-14 under HEADAS 6.0). We applied standard screening criteria to
the data (CCD temperature $T<-45\deg$ C, eliminated hot pixels and bad
aspect times). Hot and flickering pixels were removed. High background
intervals due to dark current enhancements and bright Earth limb were
removed too. Screened event files where then used to derive light
curves and spectra. We included data between 0.5 and 10 keV, where PC
response matrix is calibrated (we used the v.8 response
matrices).

We extracted data from two WT observations. The extraction region is
computed automatically by the analysis software and is a box 40 pixels
along the WT strip, centered on source, encompassing $\sim 98\%$ of the
Point Spread Function in this observing mode. We extracted photons
from PC data from an annular region (3 pixels inner radius, 30 pixel
outer radius) in order to avoid pile-up contamination. We consider
standard grades 0--2 in WT and 0--12 in PC modes. Background spectra
were taken from close-by regions free of sources.

\subsection{Timing Analysis}
\label{time}

Data were barycentered using the FTOOL task {\tt barycorr} to correct
the photon arrival times to the Solar system barycenter. A period
search led to a clear detection of the neutron star spin period. The
best period is $P=11.0027\pm0.0003$ s (all errors in the text are
given at $90\%$ confidence level). This has been derived with phase
fitting techniques. This period is consistent with the extrapolation from
known ephemerides at a constant period derivative (Kaspi \& Gavriil
2003; Dall'Osso et al. 2003).

We divided the data in four energy bands then folded them at the neutron star 
spin period. The pulse profiles were then fitted
with a sine wave obtaining the pulsed fraction (PF) in different energy ranges.
We define here as PF the (semi-)amplitude of the best fitting sine to 
the normalised and background corrected folded data. We found a PF of
$31\pm2\%$, $39\pm3\%$, $29\pm4\%$ and $35\pm7\%$ in the 0.2--10 keV, 0.2--2
keV, 2--4 keV and 4--10 keV energy bands, respectively.

%%%%%%%%%%%%%%%%%%  TABLE SPECTRA %%%%%%%%%%%%%%%%%%

\begin{table*}[t]
\begin{center}
\caption{Spectral parameters of the {\em Chandra} ACIS--S observation (4.5
count s$^{-1}$) and of the simultaneous fits of the 5 longer {\em Swift}
observations (0.3 count s$^{-1}$ on average). Fluxes are in the 0.5-10\,keV
band in units of $10^{-10}$\,erg\,s$^{-1}$\,cm$^{-2}$; column density were
fixed at the {\em XMM--Newton} value of 1.36$\times 10^{22}$\,cm$^{-2}$ (Rea et
al.~2005; {\em  phabs} model in XSPEC); all errors are at $90\%$ confidence
level. Normalisations are in XSPEC units, i.e. the number of photons
kev$^{-1}$ s$^{-1}$ cm$^{-2}$ at 1 keV.}
{\small 
\begin{tabular}{lcccc} 
\hline
 & \multicolumn{2}{c}{{\em Chandra} ACIS--S} & \multicolumn{2}{c}{{\em Swift}--XRT} \\

\hline
 & PL & PL+BB & PL & PL+BB \\
\hline

$\Gamma$ & $3.11^{+0.02}_{-0.02}$   & $2.74^{+0.02}_{-0.08}$ 
                          & $2.93^{+0.04}_{-0.04}$ & $2.70^{+0.16}_{-0.19}$\\ 
Norm PL & $0.064^{+0.002}_{-0.002}$ &  $0.033^{+0.004}_{-0.004}$  
                                   & $0.056^{+0.02}_{-0.02}$ & $0.031^{+0.008}_{-0.008}$  \\
kT (keV) & -- & $0.42^{+0.02}_{-0.02}$  
                                   & -- &  $0.42^{+0.03}_{-0.02}$  \\
Norm BB & -- & $2.6^{+0.4}_{-0.6}\times 10^{-4}$  
                                   & -- & $4.7^{+1.2}_{-1.2}\times 10^{-4}$ \\
Abs. Flux & $0.40^{+0.01}_{-0.03}$ & $0.39^{+0.01}_{-0.02}$   
                                   & $0.45^{+0.01}_{-0.01}$ & $0.44^{+0.07}_{-0.06}$\\ 
Unabs. Flux & $1.31^{+0.06}_{-0.07}$ &  $1.30^{+0.07}_{-0.07}$
                                   &  $1.74^{+0.07}_{-0.07}$ &  $1.43^{+0.04}_{-0.07}$ \\
$\chi^2_{\rm red}$ (d.o.f.) & 1.76 (154) &  0.96 (152) 
                                   & 1.88 (125) & 1.00 (123) \\
\hline
\hline
\end{tabular}
}
\label{specpar}
\end{center}
\end{table*}

%%%%%%%%%%%%%%%%%%%%%%%%%%%%%%%%%%%%%%%%%%%%%%%%%%%%%%

\subsection{Spectral analysis}
\label{spec}
 
Spectral modelling was performed fitting together the five PC observations
with exposure time longer than 1 ks, grouping the PC spectra to 60 counts per
energy bin. The spectra were fitted in the 1--10 keV energy range since the
high absorption made useless the data below 1\,keV (Romano et al.~2005). For
the PC data described above we generated the appropriate arf files with the
FTOOL task {\tt xrtmkarf} and used the latest v.8 response matrices.
The spectral parameters are reported in Table 2. 

We first fitted all the data with an absorbed (using {\tt phabs}
within XSPEC) power-law, leaving all the parameters free to vary. This
model gave a reduced $\chi^2$ value of $\chi^2_{\rm
red}=1.08$. The resulting column density is $N_{H}=(2.00^{+0.10}_{-0.16})\times
10^{22}\cmdue$ and the power law photon index is $\Gamma=3.55^{+0.08}_{-0.17}$.
We also consider a fit with the inclusion of a black body component.
In this case we derive a lower column density
$N_{H}=(1.37^{+0.27}_{-0.26})\times 10^{22}\cmdue$ consistent with the value
obtained by Rea et al. (2005). In addition we got $kT=0.42^{+0.04}_{-0.05}$ keV
and $R=5.4\pm1.5$ km (calculated at 5\,kpc distance) as well as a flatter
power law index $\Gamma=2.73^{+0.43}_{-0.52}$. Also in this case the fit is
acceptable ($\chi^2_{\rm red}=1.01$). The inclusion of the black body
component is significant at $3\,\sigma$ level (based on an F-test). 
The 0.5--10 keV absorbed (unabsorbed) flux is
$4.4^{+0.07}_{-0.06}\times 10^{-11}\ergs\cmdue$ ($1.43^{+0.87}_{-0.45}\times
10^{-10}\ergs\cmdue$), with the power law component accounting for
$72^{+7}_{-8}\%$ of the total flux. The power law component seems to be
slightly decreased from the $82\pm1\%$ for {\em XMM-Newton}. For comparison with previous
spectra we computed also the same quantities fixing the column density to the
{\em XMM-Newton} value ($N_{H, XMM}=(1.36\pm0.04)\times 10^{22}\cmdue$). Results are
reported in Table \ref{specpar} (in this case the power law contribution
amounts to $71\pm3\%$).
  
\begin{figure*}[htbp]
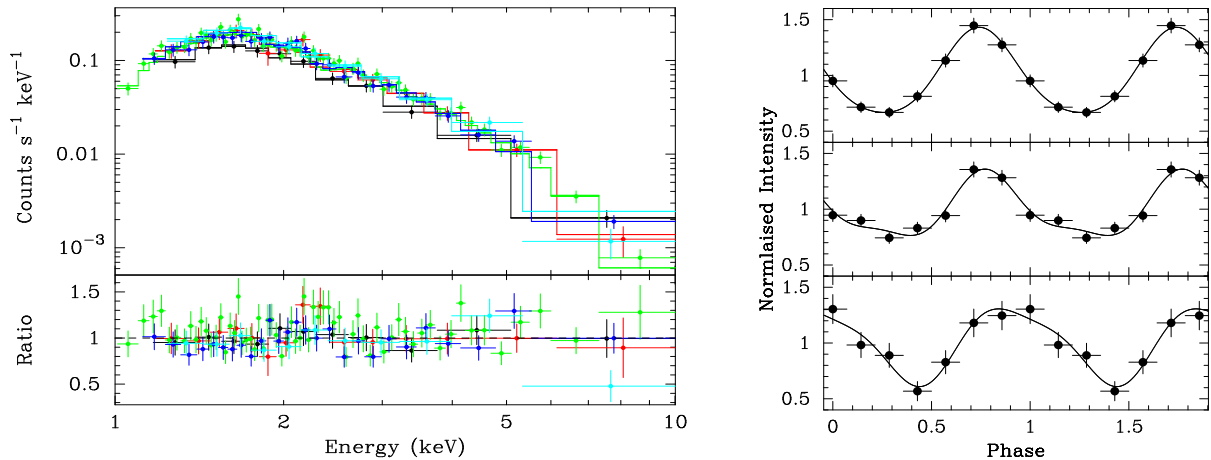

\begin{center}
\hbox{
\psfig{figure=rxs1708_spectrum.ps,height=6cm,width=9cm,angle=-90}
\hspace{0.8cm}
\psfig{figure=rxs1708_efold.ps,height=6cm,width=6cm,angle=-90}}
\caption{{\em Left Panel}: Spectra of the {\bf five longer {\em Swift} PC } observations
fitted with an absorbed blackbody plus a power-law. See Table 2 for further
details on the spectral parameters. {\em Right Panel}: folded light curves in
three energy bands, from top: 0.2--2\,keV, 2--4\,keV and 4--10\,keV.}
\end{center}
\label{spe}
\end{figure*}

The \rxs\, spectrum and flux changed significantly from the
{\em XMM-Newton} observation in August 2003. Constraining all the spectral
parameters to the {\em XMM-Newton} values within their $90\%$ confidence
intervals, the resulting fit is not acceptable with $\chi^2_{\rm
red}=5.6$. In our best fit, the blackbody temperature remains
consistent with the {\em XMM-Newton} value. The photon index instead 
decreased significantly and, at the same time, the flux
increased. Interestingly, this is in good agreement with 
the correlation found in this source by Rea et al. (2005; see also below).

\section{{\em Chandra} observation}
\label{chandra}
\rxs\, was observed by {\em Chandra} on 2004 July 3 (Obs-ID: 4605), for
$\sim$30\,ks with the Advanced CCD Imaging Spectrometer
(ACIS). The ACIS CCDs S1, S2, S3, S4, I2 and I3 were on during the
observation. In order to avoid the pile-up, the source was observed in
the Continuous Clocking (CC) mode (CC33\_FAINT; time resolution
2.85\,ms). The source was positioned in the back-illuminated ACIS-S3
CCD on the nominal target position. The data were reprocessed using
CIAO software (version 3.2). 
A detailed description on the analysis procedures, such as extraction regions,
corrections and filtering applied to the source events and spectra can
be found in Rea et al.~(2005b).  

In order to perform the timing analysis we corrected the events arrival
times for the barycenter of the solar system (with the CIAO {\tt
axbary} tool) using the provided ephemeris. We used for the timing
analysis only the events in the 0.3--8\,keV energy range and the
standard {\it Xronos} tools (version 5.19). One fundamental peak plus
one harmonic were present in the power-spectrum. A period of
$11.00223\pm0.00005$ s has been detected referred to MJD 53189. The
pulse profile has not changed with respect to the previous detection
and the 0.3--8\,keV PF is $35.4\pm0.6\%$ (see Fig. 2
right panel).

Being the CC mode not yet spectrally calibrated, the TE mode response matrices
(rmf) and ancillary files (arf) are generally used for the spectral
analysis (see Rea et al.~2005b for a detailed description
of the matrices extraction).

We fixed the absorption at $N_{H, XMM}=(1.36\pm0.04) \times 10^{22}\cmdue$
during the spectral  fitting,
because of the low statistics below 1 keV and especially because of well known 
calibration issues at 1--2\,keV, as previously reported for
other CC mode observations (Jonker et al~2003; Rea et al.~2005b).
Actually, to avoid any CC mode calibration problem, all the fits 
were performed removing the data in the 0.9--2\,keV range.

Also for this observation the best fitting model was the absorbed power law
plus a blackbody. The blackbody temperature does not change much with respect
to the {\em XMM-Newton} detection. The black body radius, $3.6\pm0.4$\,km, 
is however smaller  and the decrease appears to be significant
at or above the $3\sigma$ level.
On the other hand the power law contribution in this {\em Chandra} observation was
$80\pm2\%$, still consistent with the {\em XMM-Newton} observation of the previous
year. However, also in this case the photon index is moving harder and the
flux increasing toward what {\em Swift} saw a year later (see
Sec.\,\ref{data}). The spectral results are reported in Table 2 and
Fig. 2 (left panel). 

%%%%%%%%%%%%%%%%%%%%%%%%%%%%  FIGURE SPETTRO Chandra  %%%%%%%%%%%%%%%%%%%%%%%%%%%

\begin{figure*}[t]
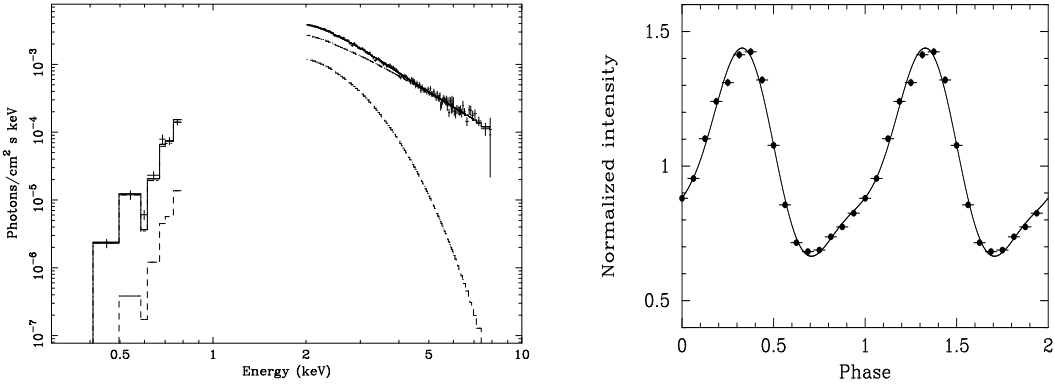

\begin{center}
\hbox{
\psfig{figure=rxs1708_cxoufs.ps,height=5cm,width=7cm,angle=-90}
\hspace{0.8cm}
\psfig{figure=rxs1708_cxoefold.ps,height=5cm,width=6cm,angle=-90}}
\caption{{\em Left Panel}: {\em Chandra} unfolded spectrum fitted with an
absorbed blackbody plus a power-law. See Table 2 for further details on the
spectral parameters. {\em Right Panel}: {\em Chandra} pulse profile in the
0.3--8.0\,keV energy band.} 
\end{center}
\label{fig_chandra}
\end{figure*}

%%%%%%%%%%%%%%%%%%%%%%%%%%%% %%%%%%%%%%%%%%%%%%%%%%%%%%%%

\section{Discussion}
\label{disc}

In this paper we present a new {\em Chandra} observation of the AXP 1RXS
J170849.0--400910 and the first {\em Swift} observations of this source
performed as a part of the XRT calibration programme.    

We performed spectral and timing analysis of the data. The measured periods,
even though affected by relatively large errors (see \S~\ref{time}),  
allows us to confirm that the sources is still in a phase of steady spin-down,
following the last glitch.

%%%%%%%%%%%%%%%%%%%%%%%%%%%%  PLOT TUTTE OSSERVAZIONI  %%%%%%%%%%
\begin{figure*}[t]
\centerline{\psfig{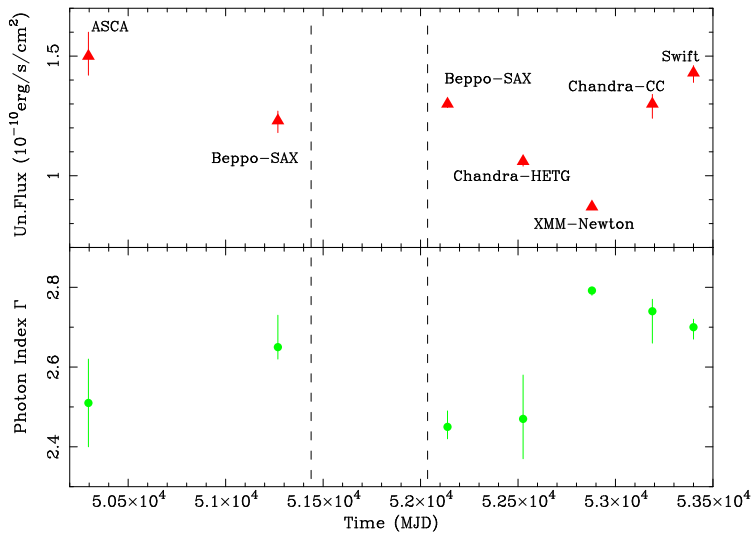}}
\caption{Long term spectral history of {\rxs}, correlated intensity-hardness
variation (adapted from Rea et al. 2005). All reported fluxes are unabsorbed
and in the 0.5--10\,keV energy range. For clarity, the observations dates are:
{\em ROSAT} -- 1994; {\em ASCA} -- 1996; first {\em BeppoSAX} -- 1997; second
{\em BeppoSAX} -- 2001;
{\em Chandra-HETG} -- 2002; {\em XMM-Newton} -- 2003; {\em Chandra-CC} -- 2004; {\em
Swift} -- 2005.}
\label{total}
\end{figure*}
%%%%%%%%%%%%%%%%%%%%%%%%%%%% %%%%%%%%%%%%%%%%%%%%%%%%%%%%

The spectral analysis reveals the source undergoing
significant spectral changes. Interestingly, the trend monitored
following the glitches epochs (Kaspi, Lackey \& Chakrabarty 2000;
Dall'Osso et al.  2003; Kaspi \& Gavriil 2003) and until the last
{\em XMM-Newton} observation has now reversed (see
Fig. \ref{total}). In particular, the source spectrum became much
harder and the total unabsorbed flux in the 0.5-10~keV energy band a
fraction $\sim 50\%$ higher with respect to that measured by {\em XMM-Newton} (Rea
et al.~2005). Moreover, our analysis indicates that the flux increase
is mainly due to an increase in the contribution of the thermal component, 
while the power law contribution to the total flux 
slightly decreased ($71\pm3\%$, while the {\em XMM-Newton}
measurement was $82\pm1\%$). 

In Rea et al. (2005) it has been proposed that the observed
correlation between the X-ray flux and the spectral hardness may be
explained within the ``twisted magnetosphere'' scenario (Thompson,
Lyutikov \& Kulkarni 2002; Beloborodov \& Thompson 2006). The basic idea is
that when a static twist is implanted, 
currents flow into the magnetosphere. As the twist angle
$\Delta\phi_\mathrm{NS}$ grows, charge carriers (electrons and
ions) provide an increasing optical depth to resonant
cyclotron scattering and hence a flatter power-law. At the same time,
the larger returning currents heat the star surface producing more
thermal photons.  Observations collected until 2003 were consistent
with a scenario in which the twist angle was steadily increasing
before the glitch epochs, culminating with glitches and a period of
increased timing noise, and then decreasing, leading to a smaller
flux and a softer spectrum.  Both the {\em Chandra} and the {\em Swift}
observations caught the source in a (relatively) hard, luminous state,
revealing a reversed trend. However, the hardening-flux correlation is
maintained, lending further support to this scenario. 

What is particularly interesting, and measured here for the first
time, is that since the last {\em XMM-Newton} observation the fraction of the 
total flux in the power-law component slightly decreased although the 
source spectrum became harder. 
This is somehow counter-intuitive. If taken at face value, it may be 
explained by the fact that, in the twisted magnetosphere model, both the 
spatial distributions of the magnetospheric currents (which act as 
``scattering medium'') and the
surface emission induced by the returning currents (which acts as
source of seed photons for the resonant scattering) are substantially
anisotropic. Seed thermal photons and scatterers are confined in two
different limited ranges of magnetic colatitudes, and both
distributions move away from the poles for larger twist angle, although
at a different rate. For instance, by using the expressions provided by
Thompson, Lyutikov \& Kulkarni (2002) for the differential luminosity
induced by the returning currents, we can estimate that the center of
the heated surface region moves from $\sim 37^\circ$ to $\sim
63^\circ$ in colatitude when $\Delta\phi_\mathrm{NS}$ increases from
$\sim 0.1$ to 2 radians.  Correspondingly, the peak of the efficiency
of the scattering only shifts from $\sim
66^\circ$ to $\sim 72^\circ$ in colatitude.  The size of the region interested
by the scattering decreases to $\sim 37\%$, while the thermally emitting
region becomes $\sim 5\%$ larger. 
Although the model is quite
approximated, and the above numbers should be treated with care, this
strong anisotropy suggests that the observed drop in the non-thermal
flux may  be due to the fact that a lower fraction of soft photons
are intercepted by the cloud of scattering particles surrounding the
star. Clearly, since the scattering depth increases with
$\Delta\phi_\mathrm{NS}$, the power-law will be in any case
flatter. 

Our preliminary quantitative estimates show, however, that
the increase in size of the thermally emitting region is not
sufficient to account for the observed variation in the blackbody
radius, at least on the basis of the original model by Thompson,
Lyutikov \& Kulkarni (2002). We only note in this respect that viewing
geometry effects may be important, since the expected change in the
position of the heated surface region may result in a larger portion
of the emitting area coming into view.

Finally, we might speculate that the long-term variations shown in
Fig. \ref{total} may have a cyclic behavior with a recurrence time
of $\approx 5$--10 yr. A possible explanation within the magnetar
scenario might be the periodic twisting/untwisting of the star
magnetosphere, where the characteristic
dissipation time of a static twist is in fact $\approx 1$--10 yr according to
the more recent estimates (Beloborodov \& Thompson 2006).  A
detailed study of this intensity-hardness correlation, through
further X-ray monitoring of this source, is needed in order to better
constrain the model, and to infer information on the physical
conditions in the star magnetosphere. Note that with a detailed
modeling of this correlation we would be able in the near future to
predict the occurrence of glitches and possibly also of
bursts. 

\begin{acknowledgements}
We thank the referee for useful comments.
SC acknowledges support from ASI (I/R/039/04 and I/023/05/0).
NR acknowledges support from an NWO Post-doctoral Fellow. 
GLI and RT acknowledge financial support from the Italian Minister of
University and Technological Research through grant PRIN2004023189. 
SZ acknowledges support from a PPARC AF.   

\end{acknowledgements}

\end{document}